\documentclass[aps,prb,twocolumn,superscriptaddress]{revtex4}

\usepackage{longtable}
\usepackage{graphicx}
\usepackage{dcolumn}
\usepackage{bm}
\usepackage{amsmath}
\usepackage[psamsfonts]{amssymb}
\usepackage{subfigure}
\usepackage{color,soul}

\newcommand{\SmB}{SmB$_6$}
\newcommand{\etal}{\textit{et al.}}
\newcommand{\MR}{magnetoresistance}

\begin{document}

\title{Signature of surface state coupling in thin films of the topological Kondo insulator {\SmB} from anisotropic magnetoresistance}

\author{M. Shaviv Petrushevsky}
\affiliation{Raymond and Beverly Sackler School of Physics and Astronomy, Tel Aviv University, Ramat Aviv, Tel Aviv 6997801, Israel}

\author{P. K. Rout}
\affiliation{Raymond and Beverly Sackler School of Physics and Astronomy, Tel Aviv University, Ramat Aviv, Tel Aviv 6997801, Israel}

\author{G. Levi}
\affiliation{Department of Materials Science and Engineering, Faculty of Engineering, Tel Aviv University, Ramat Aviv, Tel Aviv 6997801, Israel}

\author{A. Kohn}
\affiliation{Department of Materials Science and Engineering, Faculty of Engineering, Tel Aviv University, Ramat Aviv, Tel Aviv 6997801, Israel}

\author{Y. Dagan}
\email[]{yodagan@post.tau.ac.il}\affiliation{Raymond and Beverly Sackler School of Physics and Astronomy, Tel Aviv University, Ramat Aviv, Tel Aviv 6997801, Israel}

\date{\today}

\begin{abstract}
The temperature and thickness dependencies of the in-plane anisotropic magnetoresistance (AMR) of {\SmB} thin films are reported. We find that the AMR changes sign from negative ($\rho_{||}<\rho_{\perp}$) at high temperatures to positive ($\rho_{||}>\rho_{\perp}$) at low temperatures. The temperature, T$_s$, at which this sign change occurs, decreases with increasing film thickness $t$ and T$_s$ vanishes for $t$ $>$ 30 nm. We interpret our results in the framework of a competition between two components: a negative bulk contribution and a positive surface AMR.
\end{abstract}


\maketitle
\section{Introduction}
\par
In a three dimensional (3D) topological insulator, surface states with a helical Dirac dispersion result in reduced backscattering due to time reversal symmetry protection. The insulating bulk, separating opposite surfaces with reversed chirality, suppresses scattering processes between these surfaces.\cite{HasanReview,QiZhangReview} It has been theorized that topologically protected surface states can also emerge in systems where the insulating gap stems from strong electron correlations.\cite{Dzero2010,TakimotoSmB6Theory2011} One such possible system is the Kondo insulator {\SmB}.
\par
The nature of the low temperature resistivity saturation in {\SmB} has been considered a puzzle for the past 40 years.\cite{Menth1969,Allen1979,FiskPressure1994} In {\SmB}, hybridization of itinerant $\textit{d}$-electrons with localized $\textit{f}$-electrons drives the opening of a gap and surface states emerge at low temperatures.\cite{DzeroReview2016} Recent experiments\cite{SmB6CapacitanceFisk,WolfgastLowTemp,NonLocalHallFiskXia,FiskNature,PglioneThickness,HasanSmB6ARPES,XuARPESSmB6,FengSmB6ARPES,
FrantzeskakisSmB6ARPES,ZhuSmB6ARPES,PaglionePointContact,HoffmanSTM,RuanSTMSmB6} have shown strong evidence for the formation of a gap and the emergence of metallic surface states at low temperatures. Magnetotransport experiments have reported weak-antilocalization signal\cite{FiskXiaWAL} and ferromagnetism\cite{Paglione1DSmB6} below 300 mK, which have been attributed to non-trivial surface states. Quantum oscillation measurements were interpreted as evidence for these surface states;\cite{LiLudHvA} but in another experiment as a bulk signal.\cite{SebastiandHvA} A distinct evidence of the non-trivial nature of the surface states of SmB$_6$ is still lacking.
\par
In this work we measured in-plane anisotropic magnetoresistance (AMR) on thin films of {\SmB} with various thicknesses $t$. We find that the surface AMR has an opposite sign compared to that of the bulk. The surface AMR appears at a typical temperature T$_s$, which decreases with increasing $t$ and vanishes for $t$ $>$ 30 nm. The order of magnitude of T$_s$ is consistent with an excitation gap found from a model fit to the temperature dependence of the resistivity. We suggest that the surface positive AMR at low temperatures is a result of weak inter-surface coupling, which becomes insignificant as the sample becomes thicker. We use this interpretation to estimate the spin lifetime in the bulk to be of the order of 5 $\times$ 10$^{-14}$ sec.

\section{Experiment and Method}
\par
In order to maximize the surface over bulk contribution and to study the effect of inter-surface coupling as a function of thickness, we deposit thin films of {\SmB} on MgO (100) substrates using pulsed laser deposition (PLD) with a 248 nm KrF laser. One of the main difficulties in such deposition is escape of boron from the film, which leads to boron deficient {\SmB} films. Yong {\etal} have overcome this issue by co-sputtering {\SmB} and boron targets.\cite{TakeuchiThinFilms} We have employed a similar approach. To achieve the correct film stoichiometry we ablate {\SmB} and boron targets alternately with a pulse rate ratio of 3:1 respectively. The optimal growth conditions were found to be: substrate temperature of 800 $^{\circ}$C, energy density of $\sim$ 6 J$cm^{-2}$ and chamber base pressure of $4 \times 10^{-7}-1 \times 10^{-6} $ Torr. Prior to deposition the substrate is annealed for 1 hour at 900 $^{\circ}$C to remove absorbed moisture. After deposition Ar gas is introduced to the chamber for fast cool-down to prevent boron diffusion.
\begin{figure}
\begin{center}
\includegraphics[width=1\hsize]{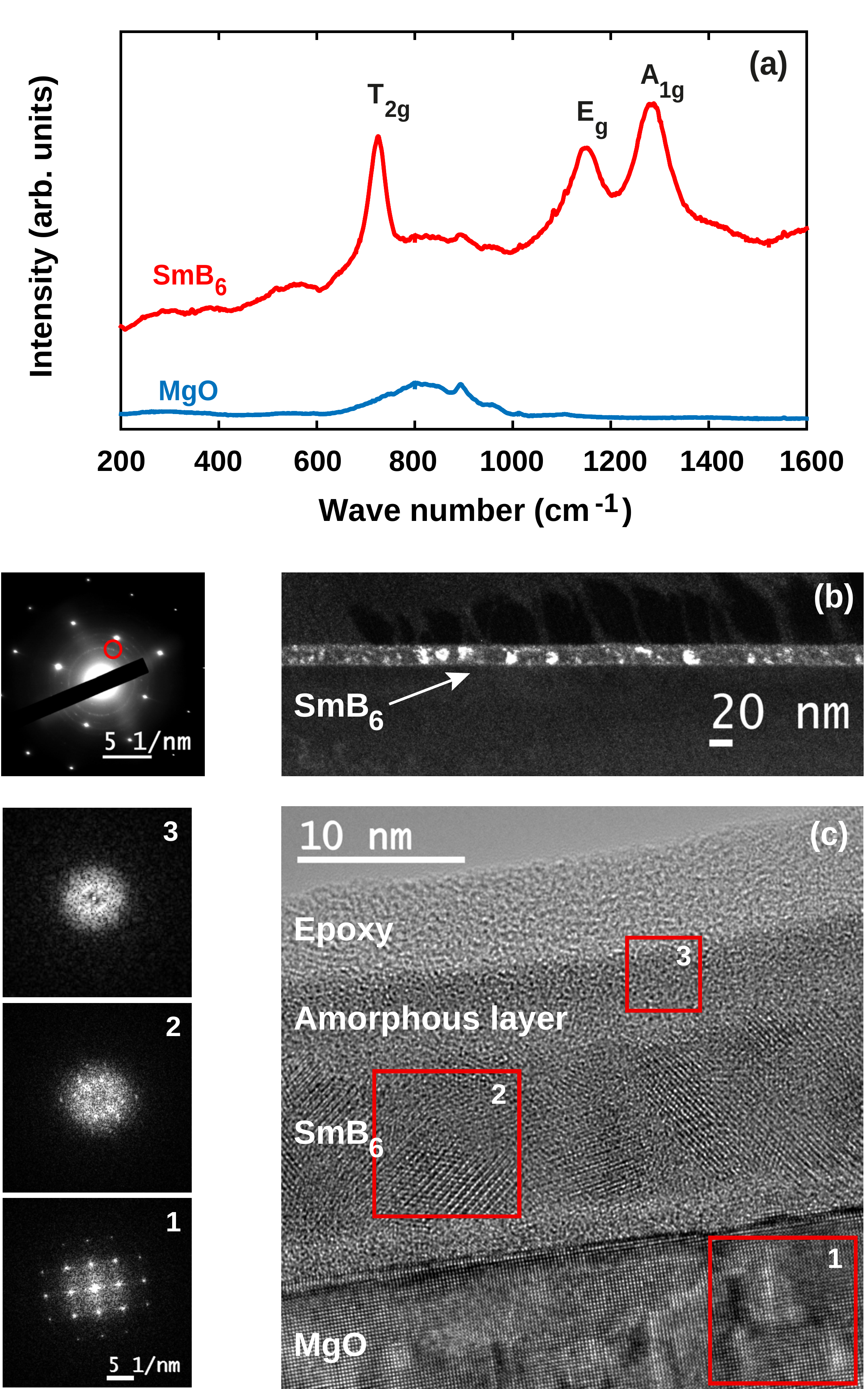}
\caption {(Color online) Characterization of 16 nm thick {\SmB} films. (a) Raman spectroscopy measurement showing the three first order Raman-allowed modes in {\SmB} due to T$_{2g}$, E$_g$ and A$_1g$ symmetries. (b) (Left) Selected area electron diffraction (SAED) of the film (cross-sectional) shows reflections from the MgO substrate in [100] zone axis and the diffraction rings from the {\SmB} film indicating a polycrystalline structure. The relative ranking of ring reflection intensities indicates random crystallographic orientation. (Right) Dark-field (DF) cross-sectional transmission electron microscopy (TEM) image constructed from a section of the (110) ring reflection, marked by the red circle in the SAED image. (c) High-resolution (HR, phase contrast) TEM (HRTEM) image reveals a polycrystalline cubic structure with random orientations. On the left of the HRTEM image, typical power spectra calculated from different regions of the sample are shown. Region 1 is the MgO substrate aligned to a [100] zone axis. Region 2 shows an example of an {\SmB} crystal in zone axis (also [100] zone axis). Due to the random crystallographic orientation of the film and the small crystallographic domain size compared to the film thickness, reflections from additional crystals not in zone axis can also be observed. At the {\SmB}/MgO interface, the layer is of reduced crystallinity. The top part of the film (region 3) is oxidized and amorphous.
  \label{fig:MaterialCharacterization}}
\end{center}
\end{figure}

\par
Thin film growth parameters were optimized using X-ray photoelectron spectroscopy measurements from which film stoichiometry was inferred and found to remain constant throughout the film thickness (see Appendix A). The film composition was also verified by energy dispersive X-ray spectroscopy and time-of-flight secondary ion mass spectrometry (ToF-SIMS) measurements, not shown here. Raman spectra measured with a 488 nm laser (Fig. \ref{fig:MaterialCharacterization} (a)) confirms the presence of first order Raman-allowed modes in {\SmB} - T$_{2g}$, E$_g$ and A$_{1g}$ at 730, 1100 and 1200 cm$^{-1}$ respectively \cite{SmB6Raman}. To check the crystalline quality of our films, we have performed transmission electron microscopy (TEM) measurements, which revealed polycrystalline cubic structure of {\SmB} with the Pm-3m (221) symmetry (see Fig. \ref{fig:MaterialCharacterization}(b)-(c)). The dark field TEM image shows that crystallographic domain size ranges from several nanometers to around 10 nm, which is smaller than the film thickness. A layer of reduced crystallinity at the interface between the MgO substrate and {\SmB} film is observed. The surface region of the film is of an amorphous structure due to oxidation as measured by energy-filtered TEM, possibly created during sample preparation for TEM measurements. Energy-filtered TEM measurements also confirm the correct stoichiometry and homogeneity of the film within error.
\par
\begin{figure}
\begin{center}
\includegraphics[width=1\hsize]{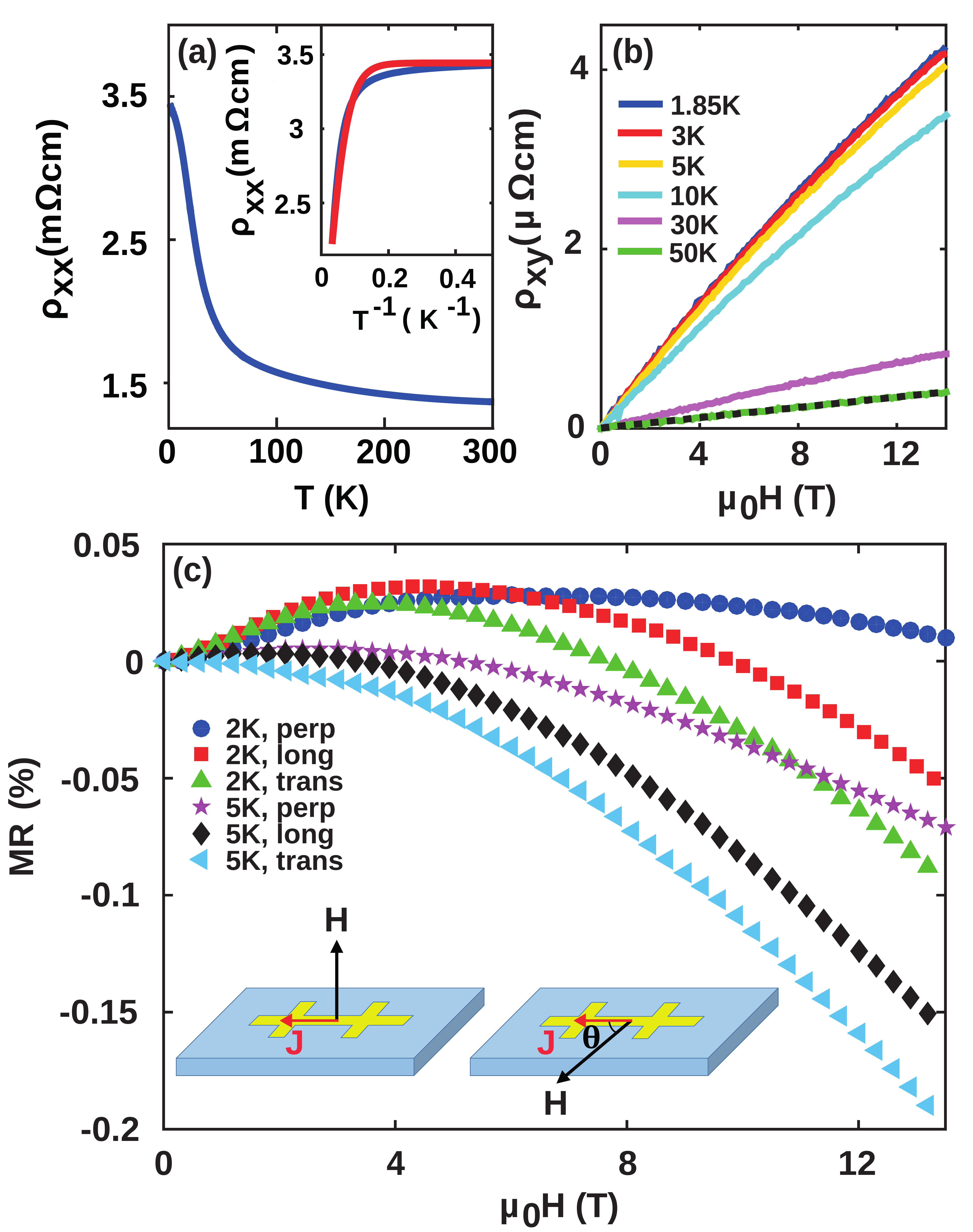}
\caption {(Color online) Transport measurements on a 16 nm thick Hall bar sample. (a) The typical longitudinal resistivity as a function of temperature. The inset shows the fit (red line) to the data below 30 K according to parallel conduction model. (b) Hall resistivity at different temperatures. The dashed black line is linear fit to the data at 50 K. (c) Magnetoresistance,  $MR = \frac{\rho_{xx}(\mu_0 H)}{\rho_{xx}(0)}-1$, at 2 K and 5 K. The inset depicts the measurement configurations for both perpendicular (perp) and in-plane magnetic field measurements: longitudinal (long, $\theta=0^\circ$) and transverse (trans, $\theta=90^\circ$). Measurements presented in (b) are with the perpendicular configuration. \label{fig:BasicTransport}}
\end{center}
\end{figure}

For transport measurements thin films with thickness ranging from 12 nm to 32 nm were used. For most samples, a precision diamond scriber was used to scribe 0.5 mm wide stripes to define current flow geometry that allowed simultaneous measurement of resistivity and Hall signals. For the 16 nm sample we also used photolithography with ion-milling to define a 100 $\mu$m $\times$ 260 $\mu m$ Hall bar. Patterned and scribed samples gave qualitatively the same results. The magnetic field and current geometries are depicted in the inset of Fig. \ref{fig:BasicTransport}(c). Standard definitions for $\rho_{xx}$ and $\rho_{xy}$ are used. We took extra care to isolate the in-plane field effects from spurious effects caused by possible substrate misalignment or stage wobble. We always measured two devices simultaneously with a well defined angle between them (45$^\circ$ or 90$^\circ$). All effects reported here have the correct phase shift for the two corresponding bridges. We also measured in both positive and negative magnetic fields and symmetrized the data to eliminate spurious Hall contributions.

\section{Results and Discussion}
\par
In Fig. \ref{fig:BasicTransport}(a) we present typical longitudinal resistivity ($\rho_{xx}$) versus temperature $T$ data measured in the 16 nm thick Hall bar sample. The observed behavior is similar to previous reports and has been related to a Kondo insulating bulk with conducting surfaces.\cite{PglioneThickness, WolfgastLowTemp} Taking into account the small sample thickness we find the resistance ratio (R(2 K)/R(300 K)) to be consistent with previous reports.\cite{PglioneThickness, WolfgastLowTemp, TakeuchiThinFilms} In the inset we fit a simplified parallel conduction model,\cite{WolfgastLowTemp, PglioneThickness} where the total sample resistivity consists of independent surface and bulk contributions: $\frac{1}{\rho_{xx}}=\frac{1}{R_s\cdot t}+\frac{1}{\rho_be^{\Delta/k_BT}}$. Here, $R_s$ is the surface sheet resistance, $\rho_b$ is the bulk resistivity, k$_B$ is Boltzmann constant and $\Delta$ is the excitation gap. We assume that $R_s$ is temperature independent and obtain $\Delta=2.96\pm0.02$ meV from the fit shown in the inset of Fig. \ref{fig:BasicTransport}(a). This value does not change much with sample thickness and is consistent with other reports.\cite{WolfgastLowTemp, PglioneThickness} Further analysis of sample resistance ratio as a function of thickness is consistent with coupling between opposite surfaces for thin films (see Appendix B).
\par
The Hall resistivity for various temperatures, is shown in Fig. \ref{fig:BasicTransport}(b). At high temperatures ($T\geq30K$), $\rho_{xy}$ is linear with magnetic field indicating a single bulk channel conduction. For $T<30K$, the conduction is through both surface and bulk channels, which leads to non-linear $\rho_{xy}$. This multichannel conduction becomes weakest at the lowest measured temperature of 1.85 K, consistent with a dominant surface Hall conduction.\cite{NonLocalHallFiskXia} The magnetoresistance (MR) at 2 K and 5 K for various orientations is shown in Fig. \ref{fig:BasicTransport}(c). The overall behavior is consistent with Ref. ~\onlinecite{FiskNature}. The negative MR at high fields has been attributed to the reduction of the excitation gap by magnetic field and the liberation of bulk charge carriers.\cite{CooleyPressureGap2} More data on the temperature dependence of the MR can be found in Appendix C.
\par
Figure \ref{fig:AMR_High_Low} shows in-plane anisotropic magnetoresistance (AMR) and planar Hall effect (PHE) measurements at T = 5 K and T = 70 K for the 16 nm thick Hall bar sample.
The measurement configuration is depicted in the right inset of Fig. \ref{fig:BasicTransport}(c), where the magnetic field lies in the sample plane. We have defined the AMR signal as: $\frac{\Delta\rho_{xx}(\theta)}{\rho_{xx,avg}}=\frac{\rho_{xx}(\theta)-\rho_{xx,avg}}{\rho_{xx,avg}}$, where $\rho_{xx}(\theta)$ is the longitudinal resistivity in the presence of an in-plane magnetic field making an angle $\theta$ with the current density \textit{J}. The averaged longitudinal resistivity over full field rotation, $\rho_{xx,avg}$, comes out to be $(\rho_{||}+\rho_{\perp})/2$, where $\rho_{||}=\rho_{xx}(\theta=0^{\circ})$ is the longitudinal resistivity for $\vec{H} || \vec{J}$ and $\rho_{\perp}=\rho_{xx}(\theta=90^{\circ})$ is the transverse resistivity for $\vec{H} \perp \vec{J}$. The PHE signal defined as $\frac{\rho_{xy}(\theta)}{\rho_{xx,avg}}$, is merely a manifestation of AMR in the transverse voltage. Usually, this signal should be shifted by $45^{\circ}$ compared to the AMR signal as observed in Fig. \ref{fig:AMR_High_Low}. For a polycrystalline film, the AMR and PHE signals follow the expressions of the form:\cite{AMRPHETheory}
\begin{equation}\label{AMR_theory}
    \frac{\Delta\rho_{xx}(\theta)}{\rho_{xx,avg}}=A\cos(2\theta);~~
    \frac{\rho_{xy}(\theta)}{\rho_{xx,avg}}=A\sin(2\theta)
\end{equation}
where $A = \frac{\rho_{||}-\rho_{\perp}}{\rho_{||}+\rho_{\perp}}$ is the normalized amplitude. The measured AMR and PHE signals fit quite well to Eq. \ref{AMR_theory}, as shown in Fig. \ref{fig:AMR_High_Low}.

\par
\begin{figure}
\begin{center}
\includegraphics[width=1\hsize]{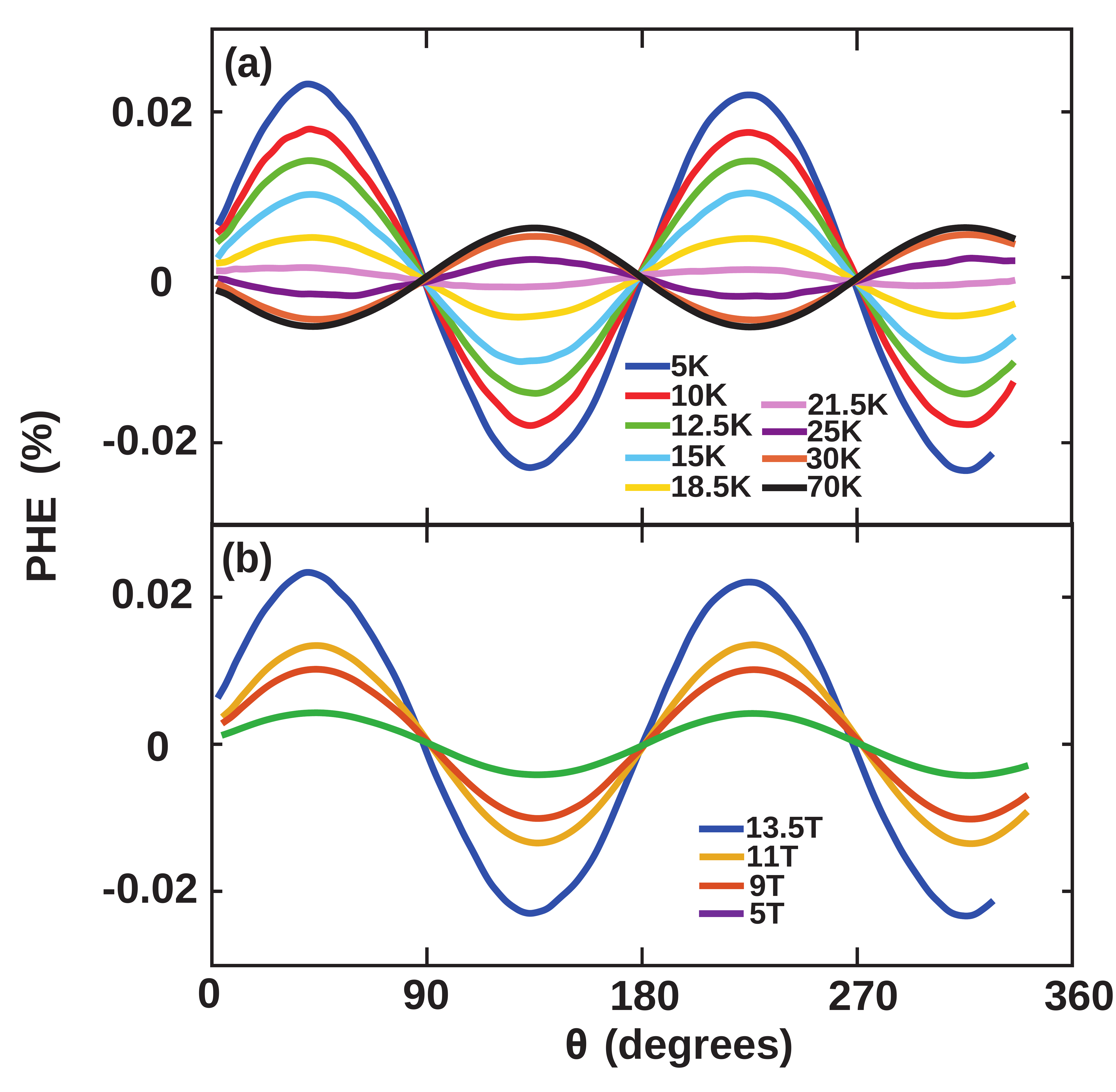}
\caption {(Color online) (a) Temperature dependence of the PHE signal for the 16 nm thick Hall bar sample measured at 13.5 T. (b) Magnetic field dependence of the PHE at 5 K for the 16 nm thick Hall bar sample. \label{fig:PHE_temp_field_dependence}}
\end{center}
\end{figure}

Surprisingly, the AMR signal has opposite signs for 5 K and 70 K. Namely, $\rho_{||}>\rho_{\perp}$ (positive AMR) at 5 K while at 70 K, $\rho_{||}<\rho_{\perp}$ (negative AMR). This sign change is observed in the PHE signal as well. We therefore carefully investigate the temperature dependence of this effect as shown in Fig. \ref{fig:PHE_temp_field_dependence}(a) for the 16 nm thick Hall bar sample. Between 70 K and 30 K, a negative signal with a negligible temperature dependence is observed. Around 23 K the AMR changes sign and as the temperature is lowered below 21.5 K, the AMR amplitude increases until it reaches a saturation value below $\simeq5 K$ (see Appendix D). The AMR amplitude decreases with decreasing field strength (see Fig. \ref{fig:PHE_temp_field_dependence}(b)), which can also be observed in Fig. \ref{fig:BasicTransport}(c).

\begin{figure}
\begin{center}
\includegraphics[width=1\hsize]{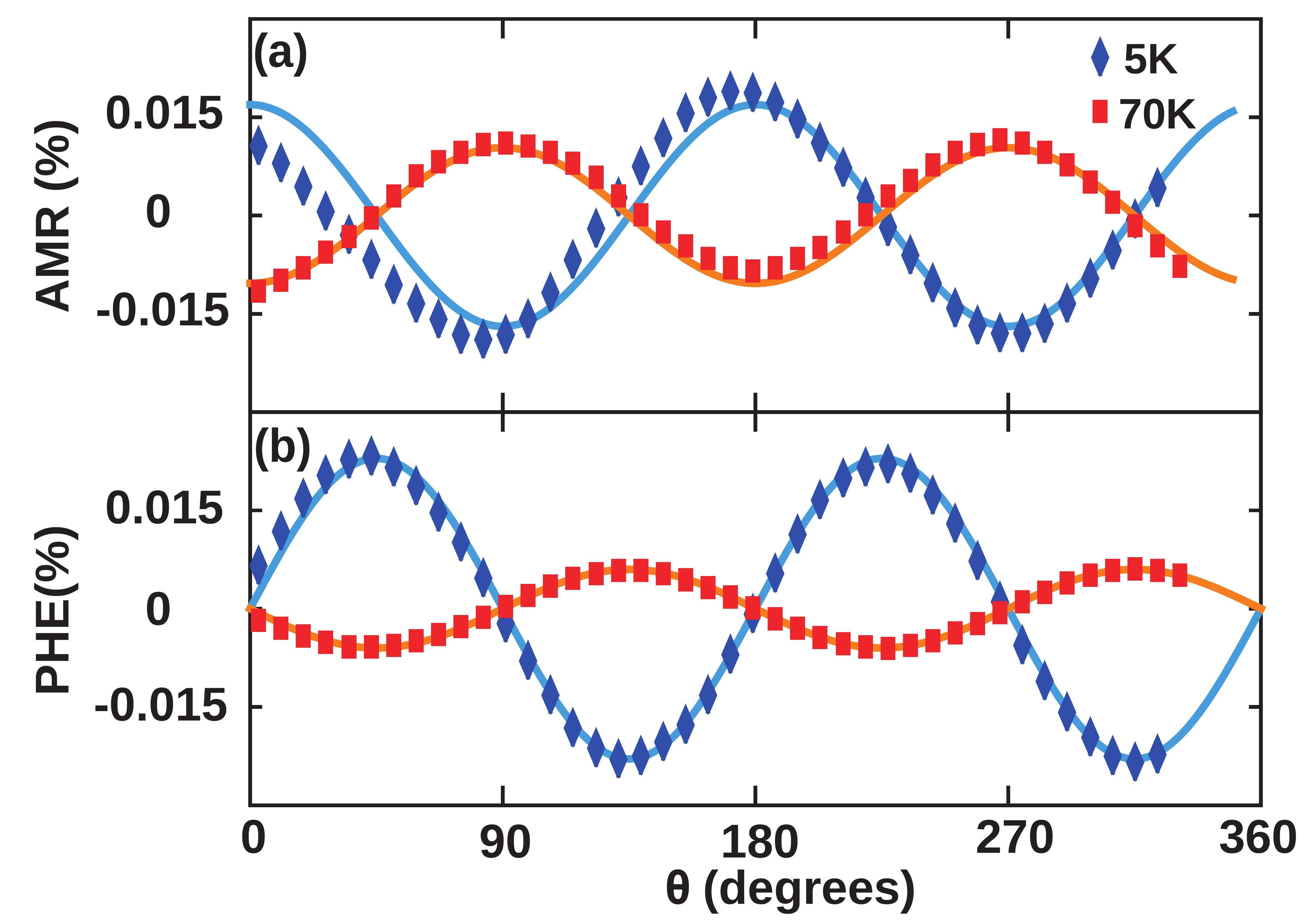}
\caption {(Color online) The AMR (a) and PHE (b) of the 16 nm thick Hall bar sample measured at magnetic field of 13.5 T and temperature of 5 K (blue diamonds) and 70 K (red rectangles). The solid lines are the fits using the theoretical expressions given in Eq. \ref{AMR_theory}. \label{fig:AMR_High_Low}}
\end{center}
\end{figure}

\par
We interpret our data as a result of a competition between surface and bulk AMR. As the temperature is lowered from 300 K to 70 K only the bulk exists, resulting in negative AMR. For 300 K $>$ T $>$ 70 K the AMR slowly increases with decreasing temperature, reaching roughly a constant value for temperatures between 70 K $>$ T $>$ 30 K (see Appendix D). At low temperatures the surface dominates over the bulk contribution and a positive AMR is seen.

\begin{figure}
\begin{center}
\includegraphics[width=1\hsize]{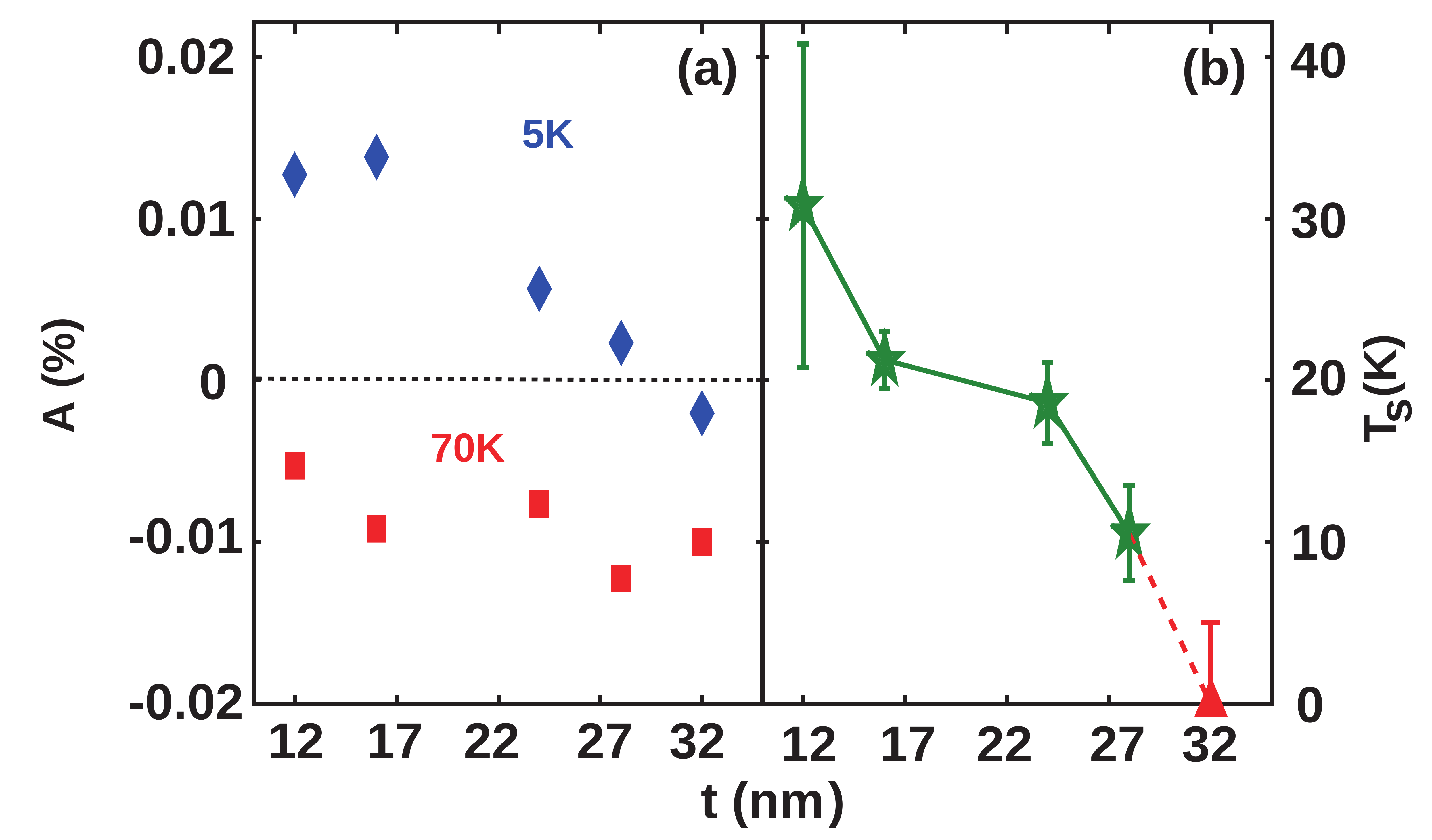}
\caption {(Color online) (a) Normalized PHE amplitude thickness dependence. The amplitude is determined by fitting the PHE data at 5 K and 70 K to Eq. \ref{AMR_theory} (blue diamonds and red rectangles, respectively). (b) Thickness dependence of the transition temperature, T$_s$, at which the PHE amplitude changes sign. For the 32 nm thick sample the PHE does not change sign, and the red triangle represents an estimation by extrapolation to low temperatures.  \label{fig:ParallelConductanceFit}}
\end{center}
\end{figure}

\par
Since we are using thin films we can study the AMR as a function of film thickness. The thickness dependence of the AMR is shown in Fig. \ref{fig:ParallelConductanceFit}(a), where we plot the normalized amplitude (extracted from fitting of the PHE signal)  at 5 K and at 70 K versus thickness, t. The normalized amplitude is roughly thickness independent at 70 K as expected for bulk effect. By contrast, the PHE amplitude decreases with increasing thickness at 5 K. We define T$_s$ as the temperature at which the PHE changes sign. We plot T$_s$ as a function of film thickness in Fig. \ref{fig:ParallelConductanceFit}(b) (see also Appendix E). Clearly, T$_s$ decreases with increasing film thickness. For our 32 nm sample we did not find a sign change down to the lowest temperature measured.
\par
The thickness dependence of the (positive) surface AMR suggests that it is related to coupling between opposite surfaces hosting states with opposite helicity. This coupling decreases with increasing thickness. The Fermi circles of these surface states can be slightly shifted due to the magnetic field \cite{AMRBSTS} or they can be different in size, e.g. due to different proximity to the substrate. We conjecture that magnetic field dependent scattering between such surfaces gives rise to the observed AMR. One can picture a mechanism somewhat similar to a two dimensional magnetic material with a Rashba-split bands as described in Ref.~\onlinecite{Jungwirth}. While in Ref.~\onlinecite{Jungwirth} the two split bands are separated only in momentum space, in our case they are also spatially separated by the bulk of the film.
\par
What is the mechanism for this inter-surface coupling? One possibility is tunneling.\cite{RashbaInducedSurfacePolarization} Another coupling mechanism is phase coherent transport through the bulk enhanced by weak anti-localization (WAL).\cite{NoteWAL,InPlaneWAL}  A third possibility is conduction via side surfaces, but this effect should be negligible for thin films. The relevant length-scale for tunneling is the wavefunction decay length, $\xi$. It has been predicted that the surface states in {\SmB} will remain topologically protected with decreasing thickness down to t $\sim$ $\xi$  $\sim$ 10 nm, due to surface Kondo breakdown.\cite{KondoBreakdown} For phase coherent transport enhanced by WAL one should consider the phase coherence length, L$_\phi$. These length scales should be compared to our film thickness.\cite{Bi2Se3Coherent,IntersurfaceCouplingBiTeSe} Recent reports estimated the dephasing length from WAL measurements in transverse magnetic fields to be of the order of 1 $\mu m$ at 20 mK, which is much greater than our film thickness.\cite{Paglione1DSmB6, FiskXiaWAL} $L_\phi$ is expected to decrease as the temperature is increased by two orders of magnitude. At 300 mK $L_\phi$ is reported to be of the order of few tens of nanometers;\cite{Paglione1DSmB6} not very far from our largest film thickness.
\par
Both coupling mechanisms described above are consistent with the observed slow rise of surface-to-bulk sheet-resistance-ratio with decreasing thickness (see Appendix B). However, for the inter-surface coupling driven AMR to work, one would require the spin to be conserved while traveling through the bulk. We can therefore estimate the spin lifetime in the bulk of our films to be of the order of 5 $\times$ 10$^{-14}$ seconds by dividing the thickness of the sample where the effect can be still observed by an average Fermi velocity taken from Ref.~\onlinecite{LiLudHvA}. Our result is longer than a naive estimation obtained from $\tau_{SO}$ = $\hbar$/E$_{SO}$, where $\hbar$ is Planck constant and E$_{SO}$ is the spin-orbit energy in {\SmB} ($\tau_{SO}$ $\sim$ 10$^{-15}$ sec).\cite{DzeroReview2016} 
\par
The absence of positive AMR for the 32 nm thick film suggests that the surfaces are decoupled down to 5 K. We explain the small negative AMR observed at this temperature as a residual bulk AMR component. This bulk contribution is an order of magnitude smaller than the surface AMR and for thinner samples it is overwhelmed by the surface contribution. Finally, we note that T$_S\simeq 30$ K determined from the AMR in our thinnest films is consistent with the temperature at which the surfaces states should be formed and with the amplitude of the excitation gap, $\Delta$.\cite{FiskNature}
\par
The TEM measurement shows that the sample used for this study has an amorphous oxidized surface. This can result from the relatively long time elapsed from sample deposition to TEM measurement. The {\SmB}/MgO interface is far from being perfect and the sample is also found to be polycrystalline with many grain boundaries. Therefore, our three channel conduction description (two surfaces with coupling and bulk) is probably an oversimplification. But this model captures the main features and it is consistent with the resistivity versus temperature and Hall data.
\section{Conclusion}
\par
We found that the in-plane anisotropic magnetoresistance (AMR) in thin films of {\SmB} changes sign with temperature. We relate the negative AMR $(\rho_\|<\rho_\bot)$ observed at high temperatures to a bulk signal, while the low temperature positive AMR is due to the surface states. The thickness dependence of the AMR and the temperature, T$_s$ at which the AMR changes sign can be explained by gradual weakening of the coupling between non-trivial surfaces with increasing thickness. Our data therefore provide evidence for the existence of nontrivial surface states in {\SmB} with a spin texture, which is distinct from the bulk. Finally, we estimate the spin life-time in the bulk to be of the order of $5\times10^{-14}$ seconds, an order of magnitude longer than the estimated spin-orbit scattering time.

\begin{acknowledgments}
We thank A. Gladkikh for the TOF-SIMS measurements, Larisa Burstein for the XPS measurements and E. Greenberg for the Raman measurements. We are indebted to M. Goldstein for helpful discussions. This work was supported in part by the Israeli Science Foundation under grant no.569/13, by the Ministry of Science and Technology under contract 3-11875 and by the Pazy foundation.
\end{acknowledgments}

\appendix
\section{X-ray Photoemission Spectroscopy measurements}
\par
\begin{figure}
\begin{center}
\includegraphics[width=1\hsize, height=0.35\textheight,keepaspectratio]{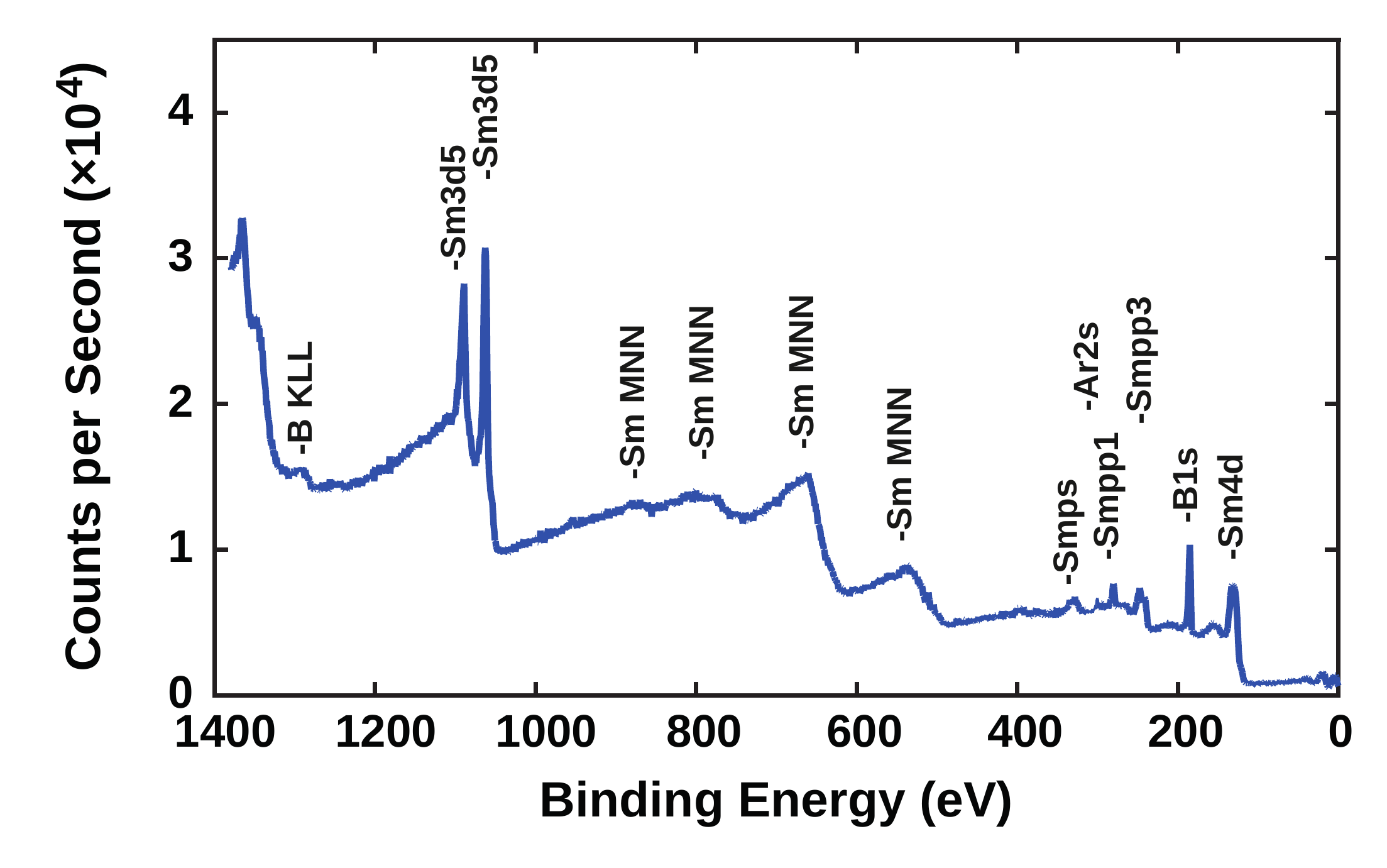}
\caption {(Color online) The XPS spectrum for a 16 nm sample after 3 minutes of sputter-cleaning. The measured region diameter was 0.8 mm.\label{fig:XPS}}
\end{center}
\end{figure}

The film growth parameters were optimized by using X-ray photoelectron spectroscopy (XPS) measurements from which the film stoichiometry was inferred. The sample was analyzed after sputter-cleaning for 1, 3 and 5 minutes using Ar$^{+}$ ion gun. In Fig. \ref{fig:XPS} we show an XPS measurement of a 16 nm sample from which we extracted the correct stoichiometric ratio of 1:6 (Sm:B). The composition was found to remain constant as a function of sputtering time, indicating a uniform composition as a function of thickness.

\section{Inter-surface coupling}
\par
\begin{figure}
\begin{center}
\includegraphics[width=1\hsize, height=0.3\textheight,keepaspectratio]{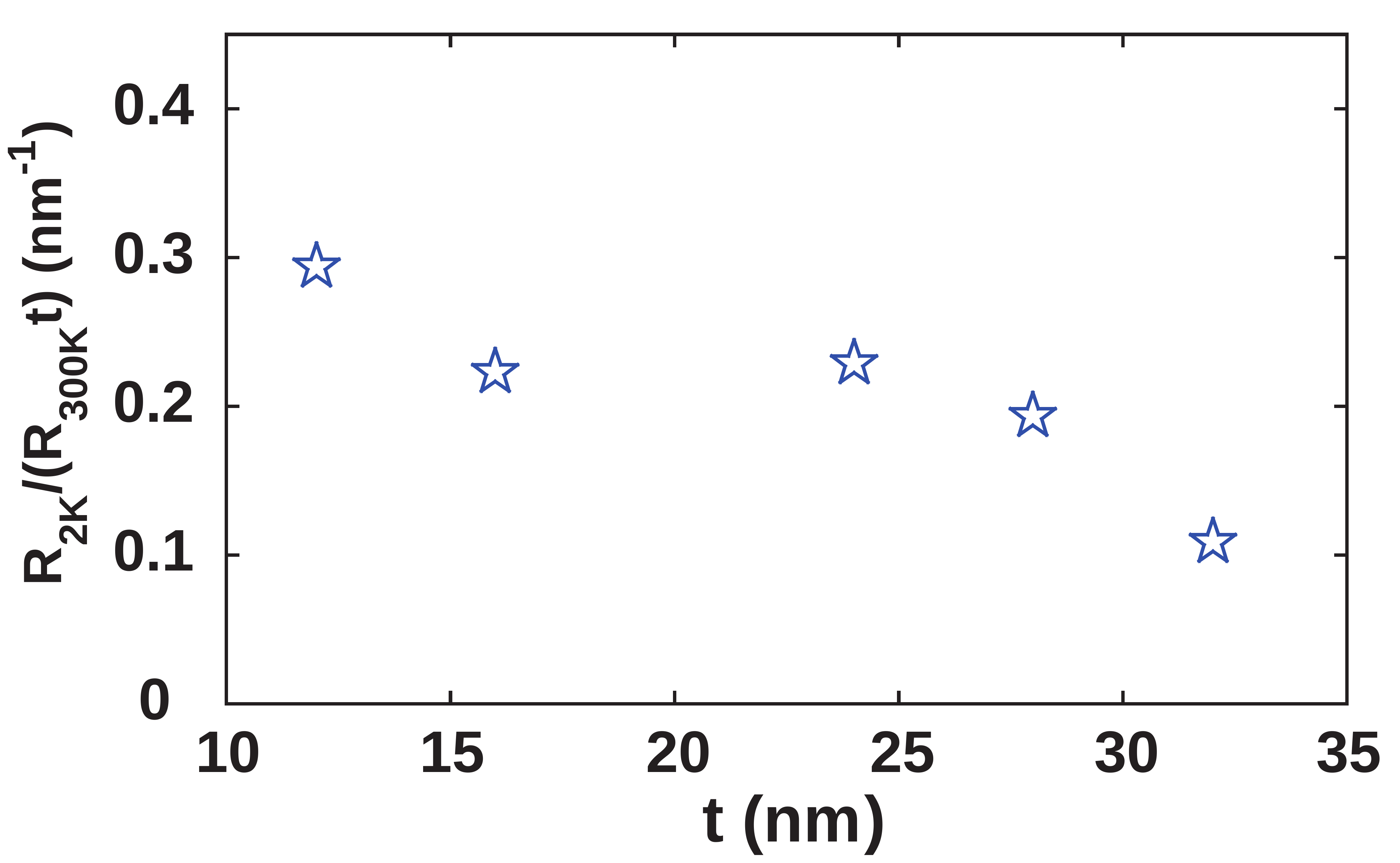}
\caption {(Color online) The resistance ratio $R_{2 K}/(R_{300 K}t)$ for different thicknesses. \label{fig:ResistanceRatio}}
\end{center}
\end{figure}

In order to study the surface contribution as a function of sample thickness, we plot the ratio $R_{2 K}/(R_{300 K}t)$ as a function of $t$ in Fig. \ref{fig:ResistanceRatio}. This method gets rid of any uncertainty arising from the sample geometry. The ratio represents the relative surface-to-bulk ratio of the resistance, which is expected to be constant in an ideal case. However, the slow rise of this ratio with decreasing thickness indicates the gradual destruction of surface states, which can happen due to the coupling between opposite surfaces for thin films.

\section{Additional Magnetoresistance Measurements}
\par
\begin{figure}
\begin{center}
\includegraphics[width=1\hsize, height=0.4\textheight,keepaspectratio]{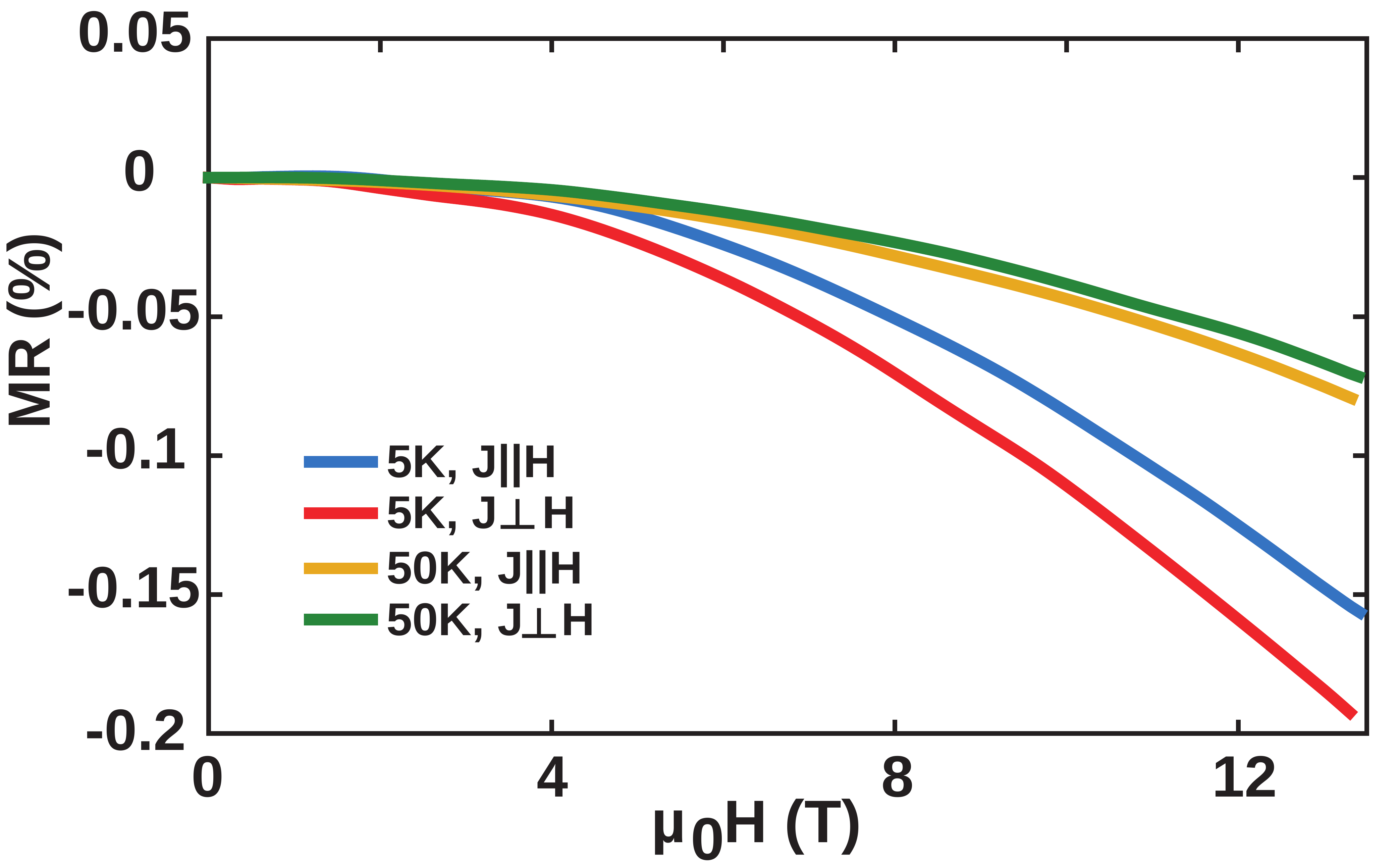}
\caption {(Color online) The in-plane magnetoresistance for the longitudinal and transverse configurations of the 16 nm Hall bar sample, showing different behaviors for the low and high temperature regimes. We can clearly see positive AMR at 5 K and negative AMR at 50 K.  \label{fig:FieldSweepsInPlaneField}}
\end{center}
\end{figure}

Figure \ref{fig:FieldSweepsInPlaneField} shows the longitudional and transverse magnetoresistance for the 16 nm Hall bar sample (discussed in the main text), which further confirms the sign change of the AMR. Comparing the magnetoresistance data at low (5 K) and high temperature (50 K), the longitudinal {\MR} is greater than the transverse one at the low temperature regime , while at the high temperature regime they are reversed.


\section{Additional anisotropic magnetoresistance and planar hall effect Measurements}

\begin{figure}
\begin{center}
\includegraphics[width=1\hsize, height=0.4\textheight,keepaspectratio]{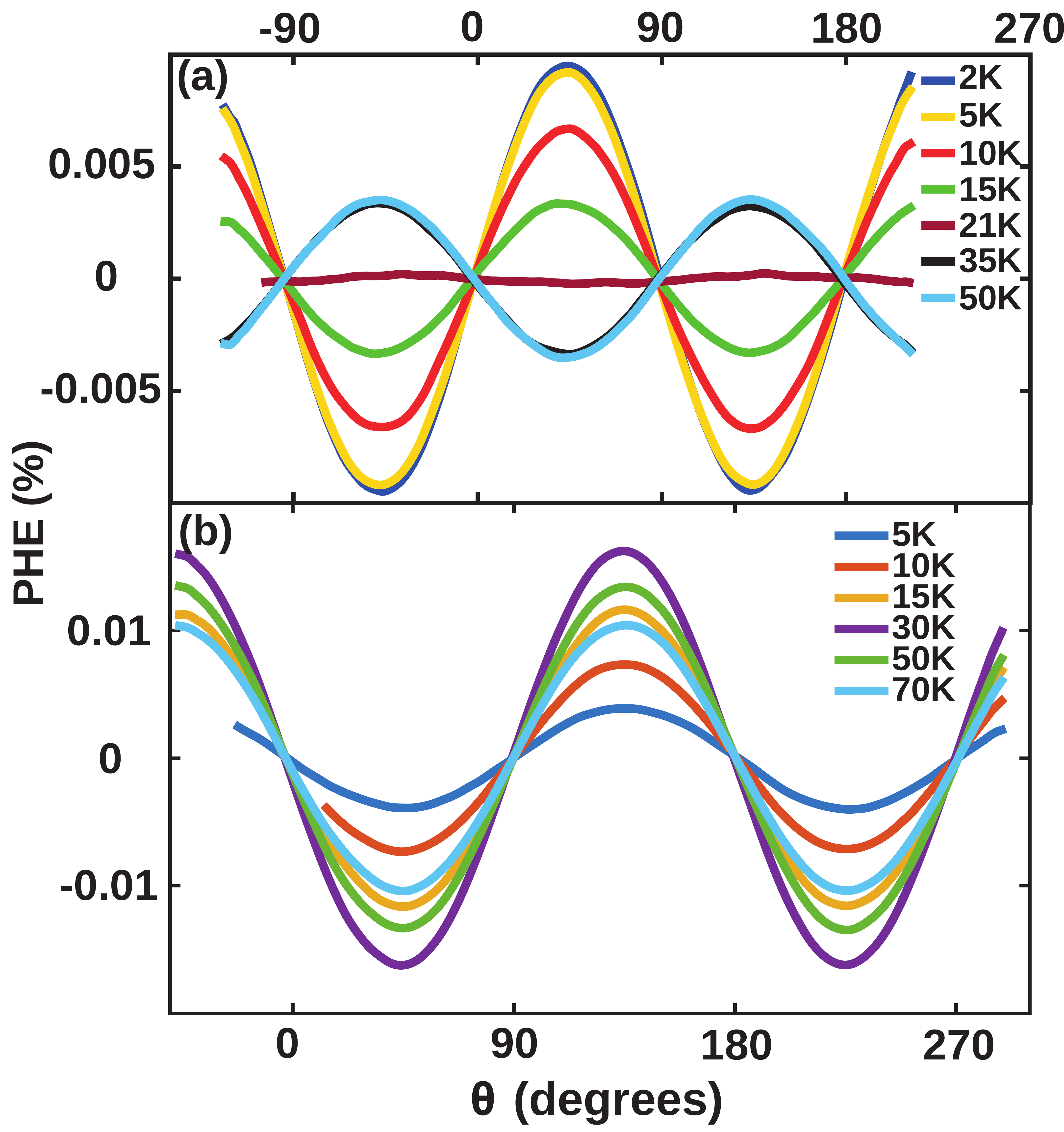}
\caption {(Color online) (a) PHE measured in a 16 nm scribed sample exhibiting low temperature saturation. (b) PHE measured in a 32 nm sample showing no transition. The PHE is always negative. \label{fig:PHE16nmScribed}}
\end{center}
\end{figure}

The temperature dependent PHE signals for 16 nm and 32 nm scribed samples are presented in Figs.  \ref{fig:PHE16nmScribed}(a) and \ref{fig:PHE16nmScribed}(b) respectively. The 16 nm scribed sample exhibits similar temperature dependence as presented in Figs. \ref{fig:AMR_High_Low} and \ref{fig:PHE_temp_field_dependence} for the 16 nm Hall bar sample. Moreover, the AMR curves for 2 and 5 K are almost identical indicating low temperature saturation of surface AMR effect. In contrast, for the 32 nm sample the AMR always remains negative and no sign change is observed with decreasing temperature.

%
\par
\begin{figure}
\begin{center}
\includegraphics[width=1\hsize, height=0.4\textheight,keepaspectratio]{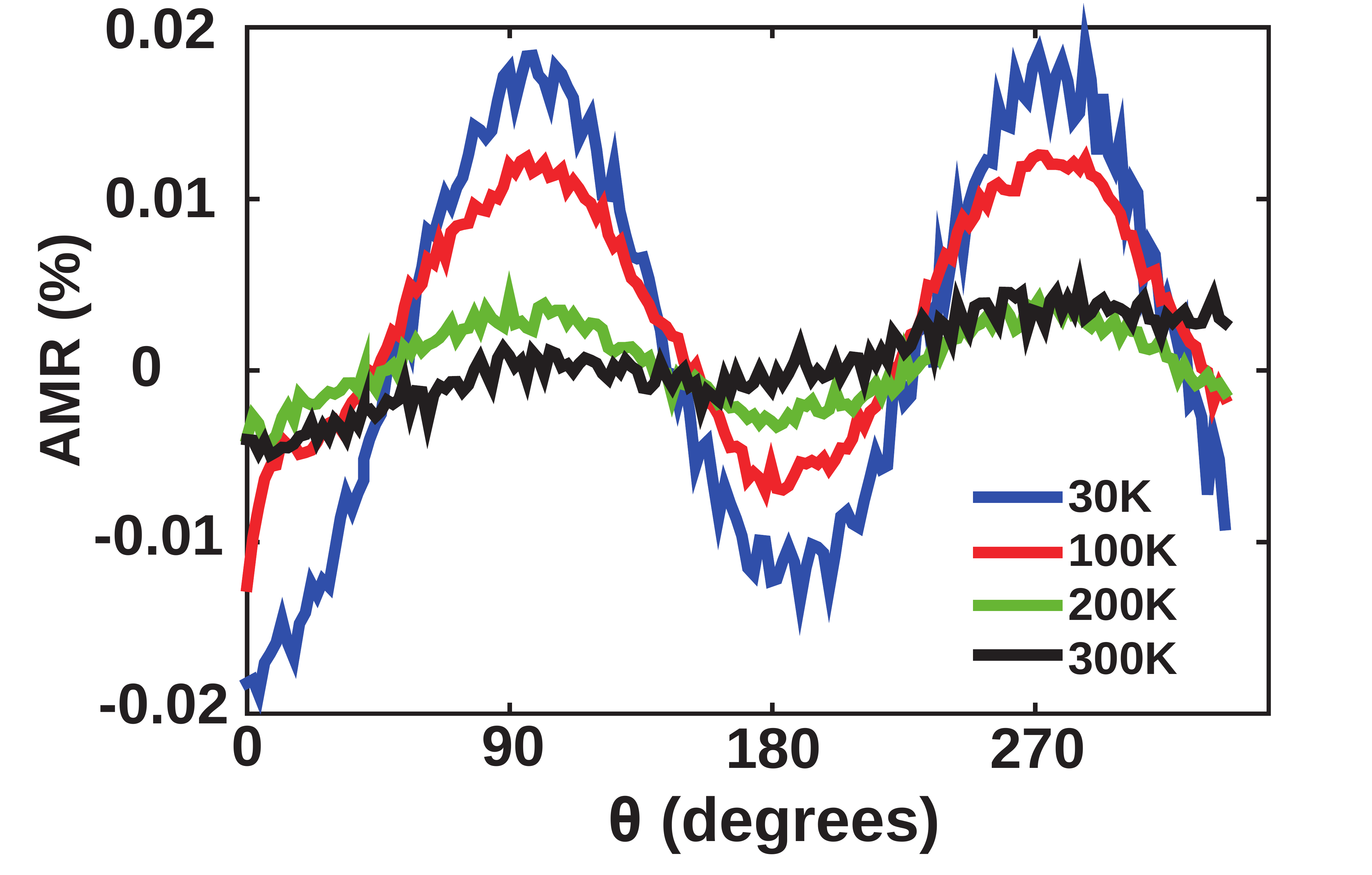}
\caption {(Color online) The AMR of the 16 nm scribed sample measured at high temperatures. The data presented here is only for positive magnetic field without any symmetrization. \label{fig:HighTempAMR}}
\end{center}
\end{figure}

Fig. \ref{fig:HighTempAMR} presents the AMR measured in a 16 nm scribed sample for high temperatures. Upon increasing the temperature above 30 K, a slow decrease in the amplitude of the bulk negative AMR  can be seen until the signal disappears between 200 K and 300 K.
%

\section{AMR transition temperature}
\par
\begin{figure}
\begin{center}
\includegraphics[width=1\hsize, height=0.35\textheight,keepaspectratio]{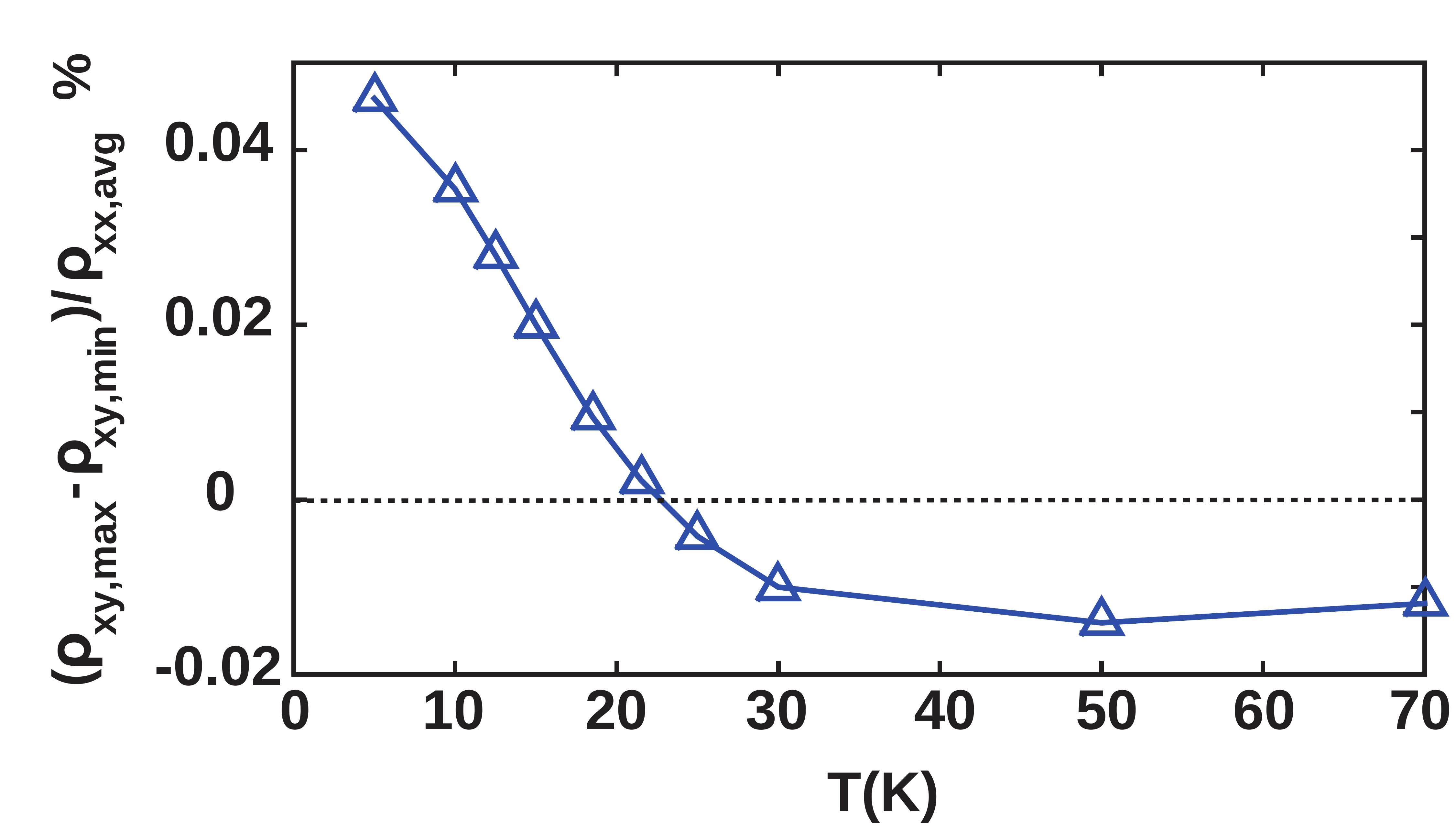}
\caption {(Color online) The measured PHE amplitude for the 16 nm thick Hall bar sample as a function of temperature. From interpolation of the curve, we estimate T$_s$ = 21$\pm1.7$ K for this sample.\label{fig:FindingTs}}
\end{center}
\end{figure}

To determine the temperature (T$_s$) at which the AMR changes sign, we have used the following method (See Fig. \ref{fig:FindingTs}). We plot the measured amplitude of the PHE as a function of the temperature and interpolate to find where the measurement curve crosses zero amplitude. The error is set as half the separation between two measurement temperatures with opposite AMR signs closest to zero amplitude.

\bibliographystyle{apsrev}
\bibliography{myBib}

\begin{thebibliography}{37}
\bibitem[{Not()}]{NoteWAL}
\bibinfo{note}{WAL in the presence of in-plane magnetic field is caused by
  finite penetration depth of the surface states into the bulk, due to the
  small excitation gap of {\SmB} \cite{Paglione1DSmB6}.}
\end{thebibliography}
\end{document}